\documentclass{ws-p8-50x6-00.new}
\usepackage{amssymb}

\newcommand{\beq}{\begin{equation}}\newcommand{\eeq}{\end{equation}}
\newcommand{\barr}{\begin{eqnarray}}\newcommand{\earr}{\end{eqnarray}}

\newcommand{\andy}[1]{ }

\newcommand{\ket}[1]{| {#1} \rangle}
\newcommand{\bra}[1]{\langle {#1} |}

\def\Ord{\mathrm{O}}
\def\As{\mathcal{A}}
\def\cH{\mathcal{H}}

\def\cR{\mathcal{R}}
\def\cZ{\mathcal{Z}}
\renewcommand{\Re}{\mathrm{Re}}
\renewcommand{\Im}{\mathrm{Im}}

\begin{document}

\title{ Unstable systems and quantum Zeno phenomena in quantum field theory}

\author{P. Facchi and S. Pascazio}

\address{Dipartimento di Fisica, Universit\`a di Bari \\
     and Istituto Nazionale di Fisica Nucleare, Sezione di Bari \\
 I-70126 Bari, Italy \\
E-mail: paolo.facchi@ba.infn.it \\
E-mail: saverio.pascazio@ba.infn.it
 }


\maketitle

\abstracts{
We analyze the Zeno phenomenon in quantum field theory. The decay
of an unstable system can be modified by changing the time
interval between successive measurements (or by varying the
coupling to an external system that plays the role of measuring
apparatus). We speak of quantum Zeno effect if the decay is slowed
and of inverse quantum Zeno (or Heraclitus) effect if it is
accelerated. The analysis of the transition between these two
regimes requires close scrutiny of the features of the interaction
Hamiltonian. We look in detail at quantum field theoretical models
of the Lee type.}

\section{Introduction }
\label{sec-introd}
\andy{intro}

The seminal formulation of the quantum Zeno effect by Misra and
Sudarshan\cite{Misra} deals with unstable systems, i.e.\ systems
that decay following an approximately exponential
law.\cite{review} Such a formulation was implicit also in previous
work,\cite{Beskow} where the features of the ``nondecay" amplitude
and probability were investigated. The attention was diverted to
oscillating systems, characterized by a finite Poincar\'e time,
when Cook published a remarkable paper,\cite{Cook88} proposing to
test the quantum Zeno effect (QZE) on a two-level system
undergoing Rabi oscillations. Although oscillating systems are
somewhat less interesting in this context, they are also much
simpler to analyze and indeed motivated interesting
experiments\cite{Itano} and lively discussions,\cite{Itanodisc}
giving rise to new ideas.\cite{Itanonew} However, interesting new
phenomena occur when one considers unstable systems, whose
Poincar\'e time is infinite:\cite{Bernardini93,FP98,Joichi98} in
particular, other regimes become possible, in which measurement
{\em accelerates} the dynamical evolution, giving rise to an {\em
inverse} quantum Zeno effect
(IZE).\cite{antiZeno,heraclitus,downconv,continuous}

The study of Zeno effects for {\em bona fide} unstable systems
requires the use of quantum field theoretical techniques and in
particular the Weisskopf-Wigner approximation\cite{Gamow28} and
the Fermi ``golden" rule:\cite{Fermi} for an unstable system the
form factors of the interaction play a fundamental role and
determine the occurrence of a Zeno or an inverse Zeno regime,
depending of the physical parameters describing the system. The
occurrence of new regimes is relevant from an experimental
perspective, in view of the beautiful experiments recently
performed by Raizen's group on the short-time non-exponential
decay (leakage through a confining potential)\cite{Wilkinson} and
on the Zeno effects for such (nonoscillating)
systems.\cite{Raizenlatest}

We analyze here the transition from Zeno to inverse Zeno in a
quantum field theoretical context, by looking in particular at the
Lee model. This is a good prototype for other field-theoretical
examples and is more general than one might
think.\cite{Peres80b,Joichi98} The usual approach to QZE and IZE
makes use of ``pulsed" observations of the quantum state. However,
one obtains essentially the same effects by performing a
``continuous" observation of the quantum state, e.g.\ by means of
an intense field, that plays the role of external, ``measuring"
apparatus. This is not a new idea,\cite{Kraus81,continuous} but
has been put on a firmer basis only recently.\cite{Schulman98} The
``continuous" formulation of the QZE has been discussed in detail
elsewhere\cite{PIO,Napoli} and will be briefly reconsidered here,
by focusing in particular on quantum field theory and its
interesting peculiarities, leading to new effects.

\section{Notation and preliminary notions: pulsed measurements}
\label{sec-dpw}
\andy{sec-dpw}

Let $H$ be the total Hamiltonian of a quantum system and $\ket{a}$
its initial state at $t=0$. The survival probability in state
$\ket{a}$ is
\andy{uno}
\beq
P(t) = |\As (t)|^2 =|\langle a|e^{-iHt}|a\rangle |^2
\label{eq:uno}
\eeq
and a short-time expansion yields a quadratic behavior
\andy{quadratic}
\beq
P(t) \sim 1 - t^2/\tau_{\rm Z}^2, \qquad \tau_{\rm Z}^{-2}
\equiv \langle a|H^2|a\rangle - \langle a|H|a\rangle^2 ,
\label{eq:quadratic}
\eeq
where $\tau_{\rm Z}$ is the Zeno time. Observe that if the
Hamiltonian is divided into a free and an interaction parts
\andy{Hdiv}
\beq
H=H_0 + H_{\mathrm{int}}, \qquad \mbox{with} \quad
H_0\ket{a}=\omega_a\ket{a}, \quad
\bra{a}H_{\mathrm{int}}\ket{a}=0,
\label{eq:Hdiv}
\eeq
the Zeno time reads
\andy{tzoff}
\beq
\tau_{\rm Z}^{-2} = \bra{a}H_{\mathrm{int}}^2\ket{a}
\label{eq:tzoff}
\eeq
and depends only on the (off-diagonal) interaction Hamiltonian.

\begin{figure}[t]
\begin{center}
\epsfig{file=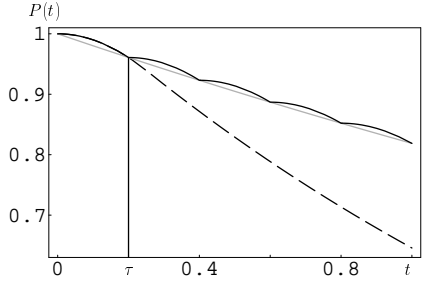,width=5.5cm}
\end{center}
\caption{Evolution with frequent ``pulsed" measurements: quantum
Zeno effect. The dashed (full) line is the survival probability
without (with) measurements. The gray line is the interpolating
exponential (\ref{eq:survN0}).}
\label{fig:zenoevol}
\end{figure}

Let us start from ``pulsed" measurements, as in the seminal
approach.\cite{Misra} The notion of ``continuous" measurement will
be discussed later. Perform $N$ (instantaneous) measurements at
time intervals $\tau=t/N$, in order to check whether the system is
still in state $\ket{a}$. The survival probability after the
measurements reads
\andy{survN}
\beq
P^{(N)}(t)=P(\tau)^N = P\left(t/N\right)^N
\sim\exp\left(-t^2/\tau_{\rm Z}^2 N\right)
\stackrel{N \rightarrow\infty}{\longrightarrow} 1 .
\eeq
If $N=\infty$ the evolution is completely hindered. For very large
(but finite) $N$ the evolution is slowed down: indeed, the
survival probability after $N$ pulsed measurements ($t=N\tau$) is
interpolated by an exponential law\cite{heraclitus}
\andy{survN0}
\beq
P^{(N)}(t)=P(\tau)^N=\exp(N\log P(\tau))= \exp(-\gamma_{\rm
eff}(\tau) t) ,
\label{eq:survN0}
\eeq
with an {\em effective decay rate}
\andy{eq:gammaeffdef}
\beq
\gamma_{\rm eff}(\tau) \equiv -\frac{1}{\tau}\log P(\tau)
= -\frac{2}{\tau}\log |\As (\tau)| =-\frac{2}{\tau}\Re [\log
\As(\tau)]
\ge0 \;
.
\label{eq:gammaeffdef}
\eeq
For $\tau\to 0 $ one gets  $P(\tau) \sim \exp (-\tau^2/\tau_{\rm
Z}^2)$, whence
\beq
\gamma_{\rm eff}(\tau) \sim \tau/\tau_{\rm Z}^2 \qquad
(\tau\to 0)
\label{eq:lingammaeff}
\eeq
is a linear function of $\tau$. Increasingly frequent measurements
tend to hinder the evolution. The {\em physical} meaning of the
mathematical expression ``$\tau\to 0$" is a subtle
issue,\cite{heraclitus,PIO} that will be touched upon also in the
present article. The Zeno evolution is represented in Figure
\ref{fig:zenoevol}.

\section{From quantum Zeno to inverse quantum Zeno (``Heraclitus")}
\label{sec-QZEinv}
\andy{QZEinv}
Let us concentrate our attention on truly unstable systems, with a
``natural" decay rate $\gamma$, given by the Fermi ``golden"
rule.\cite{Fermi} We ask: is it possible to find a finite time
$\tau^*$ such that
\andy{tstardef}
\beq
\gamma_{\rm eff}(\tau^*)=\gamma ?
\label{eq:tstardef}
\eeq
If such a time exists, then by performing measurements at time
intervals $\tau^*$ the system decays according to its ``natural"
lifetime, as if no measurements were performed. The quantity
$\tau^*$ is related to Schulman's ``jump"
time.\cite{Schulman97,review}

\begin{figure}[t]
\begin{center}
\epsfig{file=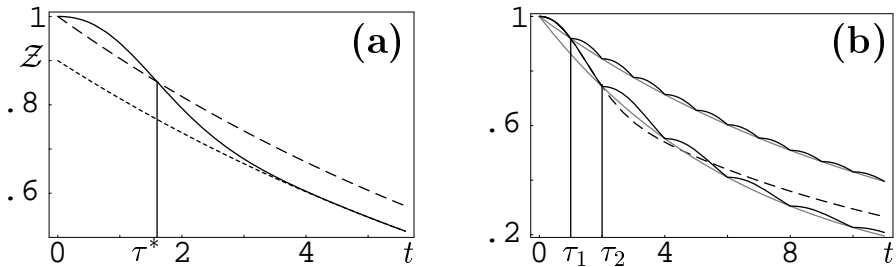,width=\textwidth}
\end{center}
\caption{$\cZ<1$. (a) Determination of $\tau^*$.
(b) Quantum Zeno vs inverse quantum Zeno (``Heraclitus") effect.}
\label{fig:gtau}
\end{figure}

By Eqs.\ (\ref{eq:tstardef}) and (\ref{eq:gammaeffdef}) one gets
\beq
P(\tau^*)=e^{-\gamma \tau^*},
\eeq
i.e., $\tau^*$ is the intersection between the curves $P(t)$ and
$e^{-\gamma t}$. In the situation outlined in Figure
\ref{fig:gtau}(a) such a time $\tau^*$ exists: the full line is
the survival probability and the dashed line the exponential
$e^{-\gamma t}$ [the dotted line is the asymptotic exponential
$\cZ e^{-\gamma t}$, that will be defined in Eq.\
(\ref{eq:psurv})]. The physical meaning of $\tau^*$ can be
understood by looking at Figure \ref{fig:gtau}(b), where the
dashed line represents a typical behavior of the survival
probability $P(t)$ when no measurement is performed: the
short-time Zeno region is followed by an approximately exponential
decay with a natural decay rate $\gamma$. When measurements are
performed at time intervals $\tau$, we get the effective decay
rate $\gamma_{\rm eff}(\tau)$. The full lines represent the
survival probabilities and the dotted lines their exponential
interpolations, according to (\ref{eq:survN0}). If
$\tau=\tau_1<\tau^*$ one obtains QZE. {\em Vice versa}, if
$\tau=\tau_2>\tau^*$, one obtains IZE. If $\tau=\tau^*$ one
recovers the natural lifetime, according to (\ref{eq:tstardef}):
in this sense, amusingly, $\tau^*$ can be viewed as a {\em
transition time} from {\em Zeno} (who argued that a sped arrow, if
observed, does not move) to {\em Heraclitus} (who replied that
everything flows). Heraclitus opposed Zeno and Parmenides' (Zeno's
master) philosophical view.\cite{PIO,Nakazato97}
\begin{figure}[t]
\begin{center}
\epsfig{file=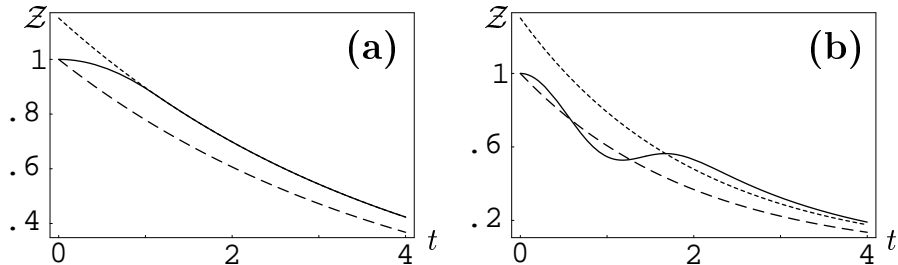,width=\textwidth}
\end{center}
\caption{$\cZ> 1$. The full line is the survival
probability, the dashed line the renormalized exponential
$e^{-\gamma t}$ and the dotted line the asymptotic exponential
$\cZ e^{-\gamma t}$. (a) If $P(t)$ and $e^{-\gamma t}$ do not
intersect, a finite solution $\tau^*$ does not exist. (b) If
$P(t)$ and $e^{-\gamma t}$ intersect, a finite solution $\tau^*$
exists. (In this case there are always at least two
intersections.) }
\label{fig:Zren}
\end{figure}

Sometimes (interestingly) $\tau^*$ does not exist: Eq.\
(\ref{eq:tstardef}) may have no finite solutions. In such a case
only QZE is possible and no IZE is attainable. This is in contrast
with some recent claims,\cite{KKclaim} according to which the
inverse Zeno regime is easier to attain than the quantum Zeno one.
We can get a qualitative idea of the meaning of these statements
by looking at the survival probability of an unstable system for
sufficiently long times\cite{review}
\andy{Psurv}
\beq
\label{eq:Psurv}
P(t)= |\As(t)|^2 \simeq  \cZ e^{-\gamma t} ,
\eeq
where $\cZ$, the intersection of the asymptotic exponential with
the $t=0$ axis, is the square modulus of the residue of the pole
of the propagator (wave function
renormalization)\cite{review,PIO,heraclitus} and will be defined
in the next section [Eq.\ (\ref{eq:psurv})].

A sufficient condition for the existence of a solution $\tau^*$ of
Eq.\ (\ref{eq:tstardef}) is that $\cZ<1$. This is easily proved by
graphical inspection. The case $\cZ<1$ is shown in Figure
\ref{fig:gtau}(a): $P(t)$ with the property (\ref{eq:Psurv}) and
$e^{-\gamma t}$ {\em must} intersect. The other case, $\cZ > 1$,
is shown in Figure \ref{fig:Zren}: a solution may or may not
exist, depending on the features of the model investigated. There
are also situations (e.g., oscillatory systems, whose Poincar\'e
time is finite) where $\gamma$ and $\cZ$ cannot be
defined.\cite{PIO} The transition from Zeno to inverse Zeno is
therefore a complex, model-dependent problem, that requires
careful investigation. We shall come back to this issue in the
following sections, where (\ref{eq:Psurv}) will be derived for a
particular field theoretical model.

\section{The Lee Hamiltonian}
\label{sec-tesi}
\andy{sec-tesi}

Some of the most interesting Zeno phenomena, including the
transition from a Zeno to a Heraclitus regime, arise in a quantum
field theoretical framework.\cite{Bernardini93,FP98,Joichi98} We
will now study the time evolution of a quantum system in greater
detail, by making use of a quantum field theoretical techniques,
and discuss the primary role played by the form factors of the
interaction.

Consider a generic Hamiltonian $H$ and an initial normalized state
$\ket{a}$. The total Hilbert space can always be decomposed into a
direct sum $\cH=\cH_{a}\oplus\cH_{d}$, with $\cH_a=P_a\cH$ and
$\cH_d=P_d\cH$, where $P_a=|a\rangle\langle a|$ and $P_d=1-P_a$.
Let us accordingly  split the total Hamiltonian into a free and an
interaction part
\beq
H=H_0+H_{\rm int},
\eeq
where
\beq
H_0=P_a H P_a + P_d H P_d, \qquad H_{\mathrm{int}}=P_a H P_d + P_d
H P_a.
\eeq
This decomposition can always be performed,\cite{Peres80b} even in
relativistic quantum field theory,\cite{Joichi98} the only
``problem" being that the decomposition itself depends on the
initial state $\ket{a}$. Let $\{\ket{n}\}$ be the eigenbasis of
$H_0$ in $\cH_d$
\barr
& &\bra{a}a\rangle=1,\quad \bra{a}n\rangle=0, \quad
\bra{n}n'\rangle=\delta_{nn'} ,
\label{eq:orthonormal}
\\
& &H_0\ket{a}=\omega_a\ket{a}, \qquad H_0\ket{n}=\omega_n\ket{n} .
\label{eq:Hamsplit}
\earr
The interaction Hamiltonian $H_{\rm int}$ is completely
off-diagonal and has nonvanishing matrix elements only between
$\cH_a$ and $\cH_d$, namely
\beq
\bra{a}H_{\rm int}\ket{n} =\bra{n}H_{\rm int}\ket{a}^*=g_n ,
\qquad \bra{a}H_{\rm int}\ket{a}=\bra{n}H_{\rm int}\ket{n'}=0,
\quad \forall n,n' .
\label{eq:intmatrel}
\eeq
Equations\ (\ref{eq:orthonormal})-(\ref{eq:intmatrel}) completely
determine the free and interaction Hamiltonians in terms of the
chosen basis. Indeed we get
\barr
H_0=\omega_a\ket{a}\bra{a}+\sum_n\omega_n\ket{n}\bra{n}, \qquad
H_{\rm int}=\sum_n \left(g_n \ket{a}\bra{n}+
g_n^*\ket{n}\bra{a}\right).
\label{eq:Lee}
\earr
This is the Lee Hamiltonian\cite{Lee54} and was originally
introduced as a solvable quantum field model for the study of the
renormalization problem. The interaction of the normalizable state
$\ket{a}$ with the states $ \ket{n}$ (the formal sum in the above
equation usually represents an integral over a continuum of
states)  is responsible for its decay and depends on the
\emph{form factor} $g_n$.

The Fourier-Laplace transform of the survival amplitude $\As(t)$
in (\ref{eq:uno}) is the expectation value of the resolvent
\andy{transf}
\barr
G_a(E)&=&-i \int_0^{\infty}dt\;e^{iEt}\As(t)
=\bra{a}\frac{1}{E-H}\ket{a}, \nonumber\\
\As(t)&=&\frac{i}{2\pi} \int_{\rm B}dE\;e^{-iEt}G_a(E) ,
\label{eq:transf}
\earr
the Bromwich path B being a horizontal line $\Im E=$constant$>0$
in the half plane of convergence of the Fourier-Laplace transform
(upper half plane). By performing Dyson's resummation, the
propagator reads
\andy{propag}
\beq
\label{eq:propag}
G_a(E)=\frac{1}{E-\omega_a-\Sigma_a(E)},
\eeq
where the self-energy function
\andy{selfen}
\beq
\label{eq:selfen}
\Sigma_a(E)=\sum_n\frac{\left|\bra{a}H_{\rm
int}\ket{n}\right|^2}{E-\omega_n}=\int
d\omega\;\frac{\kappa_a(\omega)}{E-\omega}
\eeq
consists only of a second order contribution and is related to the
form factor $g_n$ by the equation
\beq
\kappa_a(E)=\sum_n\left|\bra{a}H_{\rm
int}\ket{n}\right|^2\delta(E-\omega_n)=\sum_n
|g_n|^2\delta(E-\omega_n).
\eeq
A comment is now in order. If one is only interested in the
survival amplitude [or, equivalently, in the expression of the
propagator\ (\ref{eq:propag})] and not in the details of the
interactions $g_n$ between $\ket{a}$ and different states
$\ket{n}$ with the same energy $\omega_n=\omega$, one can simply
replace this set of states with a single, representative state
$\ket{\omega}$, by replacing the Hamiltonian (\ref{eq:Lee}) with
the following equivalent one
\andy{contham}
\beq
H=H_0+H_{\rm int} = \omega_a\ket{a}\bra{a} +\int d\omega\;\omega
\ket{\omega} \bra{\omega} +\int d\omega\;g(\omega) ( \ket{a}
\bra{\omega}+ \ket{\omega} \bra{a}) ,
\label{eq:contham}
\eeq
where the form factor $g(\omega)=\sqrt{\kappa_a(\omega)}$ and
\andy{idres}
\beq
\label{eq:idres}
\ket{a}\bra{a}+\int d\omega\;\ket{\omega}\bra{\omega}= 1.
\eeq
In terms of the Hamiltonian\ (\ref{eq:contham}) the self-energy
function simply reads
\beq
\Sigma_a(E)= \int d\omega\; \frac{\left|\bra{a}H_{\rm
int}\ket{\omega}\right|^2}{E-\omega} =\int
d\omega\;\frac{g^2(\omega)}{E-\omega}  .
\label{eq:selfencont}
\eeq

\section{Unstable systems}
\label{sec-unstable}

We consider now the case of an unstable system. The initial state
has energy $\omega_a>\omega_g$ ($\omega_g$ being the lower bound
of the continuous spectrum of the Hamiltonian $H$) and is
therefore embedded in the continuous spectrum of $H$. If
$-\Sigma_a(\omega_g)<\omega_a$ (which happens for sufficiently
smooth form factors and small coupling), the resolvent is analytic
in the whole complex plane cut along the real axis (continuous
spectrum of $H$). On the other hand, there exists a pole
$E_{\mathrm{pole}}$ located just below the branch cut in the
second Riemann sheet, solution of the equation
\beq
E_{\rm pole}-\omega_a-\Sigma_{a{\rm II}}(E_{\rm pole})=0,
\label{eq:1equpol}
\eeq
$\Sigma_{a{\rm II}}$ being the determination of the self-energy
function in the second sheet. The pole has a real and imaginary
part
\beq
E_{\rm pole}=\omega_a + \delta\omega_a-i\gamma/2
\eeq
given by
\barr
\delta\omega_a&=&\Re \Sigma_{a{\rm II}}(E_{\rm pole})\simeq \Re
\Sigma_a(\omega_a+i0^+)={\rm P}\!\!\int d\omega
\frac{g^2(\omega)}{\omega_a-\omega} ,
\label{eq:2shift}
\\
\gamma&=&-2\Im \Sigma_{a{\rm II}}(E_{\rm pole})\simeq -2\Im
\Sigma_a(\omega_a+i0^+)=2\pi g^2(\omega_a) ,
\label{eq:FGR}
\earr
up to fourth order in the coupling constant. One recognizes the
second-order energy shift $\delta\omega_a$ and the celebrated
Fermi ``golden" rule $\gamma$.\cite{Fermi} The survival amplitude
has the general form
\beq
\As(t)=\As_{\rm pole}(t)+\As_{\rm cut}(t),
\eeq
where
\beq
\As_{\rm pole}(t)=e^{-i(\omega_a+\delta\omega_a)t-\gamma
t/2}/[1-\Sigma'_{a{\rm II}}(E_{\rm pole})],
\eeq
$\As_{\rm cut}$ being the branch-cut contribution. At intermediate
times, the pole contribution dominates the evolution and
\andy{psurv}
\beq
P(t)\simeq |\As_{\rm pole}(t)|^2 =  \cZ e^{-\gamma t} ,\qquad
\cZ=\left|1-\Sigma'_{a{\rm II}}(E_{\rm pole})\right|^{-2} ,
\label{eq:psurv}
\eeq
where $\cZ$, the intersection of the asymptotic exponential with
the $t=0$ axis, is the wave function renormalization. We have
found (\ref{eq:Psurv}) and explicitly determined $\cZ$.

Notice that, in order to obtain a purely exponential decay, one
neglects all branch cut and/or other contributions from distant
poles and considers only the contribution of the dominant pole. In
other words, one does not look at the rich analytical structure of
the propagator and retains only its dominant polar singularity. In
this case the self-energy function becomes a constant (equal to
its value at the pole), namely
\andy{WW0}
\beq
G_a(E)
\longrightarrow G_a^{\rm
WW}(E)= \frac{1}{E-\omega_a-\Sigma_{a\rm II}(E_{\rm pole})}
=\frac{1}{E-E_{\rm pole}},
\label{eq:WW0}
\eeq
where in the last equality we used the pole equation\
(\ref{eq:1equpol}). This is the celebrated Weisskopf-Wigner
approximation\cite{Gamow28} and yields a purely exponential
behavior, $\As(t)=\exp(-iE_{\rm pole} t)$, without short- and
long-time corrections.

As is well known, the exponential law is corrected by the cut
contribution, which is responsible for a quadratic behavior at
short times and a power law at long times. In particular, at short
times, by plugging (\ref{eq:propag}) into (\ref{eq:transf}) and
changing the integration variable $\eta=Et$, Eq.\
(\ref{eq:transf}) becomes
\andy{Laplace}
\beq
  \As(t)={i\over2\pi }
  \int_B d\eta\;\frac{e^{-i\eta}}
  {\eta-\omega_a t-t\;\Sigma_a(\eta/t)} .
\label{eq:Laplace}
\eeq
The self-energy function (\ref{eq:selfencont}) has the following
behavior at large energies
\andy{Sigma(1/u)1}
\beq
\Sigma_a\left(E\right)\sim\frac{1}{E}\int d\omega g^2(\omega)=
\frac{1}{E}\bra{a}H_{\mathrm{int}}^2\ket{a}=\frac{1}{\tau_{\mathrm{Z}}^2E},
\qquad E\to\infty
\label{eq:Sigma(1/u)1}
\eeq
where we used Eq.\ (\ref{eq:tzoff}) (and assumed the existence of
the second moment of the Hamiltonian $H_{\mathrm{int}}$).
Therefore, the survival amplitude at small times has the
asymptotic expansion
\andy{aUasmallt}
\barr
\As(t) \sim \frac{i}{2 \pi} \int_B d\eta \, \frac{\eta e^{-i\eta}}
{\eta^2 - \omega_a t \eta - t^2/\tau_{\rm Z}^2} =\frac{i}{2 \pi }
\int_B d\eta \, \frac{\eta
e^{-i\eta}}{(\eta-t\eta_1)(\eta-t\eta_2)},
 \label{eq:aUasmallt}
 \earr
where
 \andy{sigma12}
 \beq
 \eta_{1,2} = \frac{\omega_a}{2} \pm \sqrt{\left(\frac{\omega_a}{2}\right)
^2+\frac{1}{\tau_{\rm Z}^2}}.
 \label{eq:sigma12}
 \eeq
By closing the Bromwich path in Eq.\ (\ref{eq:aUasmallt}) with a
large semicircle in the lower half plane, the integral reduces to
the sum of the residues at the real poles $t\eta_{1,2}$ and the
survival probability at small times reads
\andy{Psmallt2}
\barr
P(t)=|\As(t)|^2 \sim
\frac{\eta_1^2+\eta_2^2-2\eta_1\eta_2\cos[t(\eta_1-\eta_2)]}
{(\eta_1-\eta_2)^2}
\sim 1 + \eta_1 \eta_2 t ^2 =1 - \frac{t^2}{\tau_{\rm Z}^2}, \ \
\label{eq:Psmallt2}
\earr
in agreement with the expansion\ (\ref{eq:quadratic}). Notice that
at short times the behavior is governed by two ``effective" poles
which replace the global contribution of the cut and the pole on
the second sheet. We will come back to this important point in the
following sections.

\section{Two-pole model and two-pole reduction}
\label{sec-two-pole}

\begin{figure}[t]
\centerline{\epsfig{file=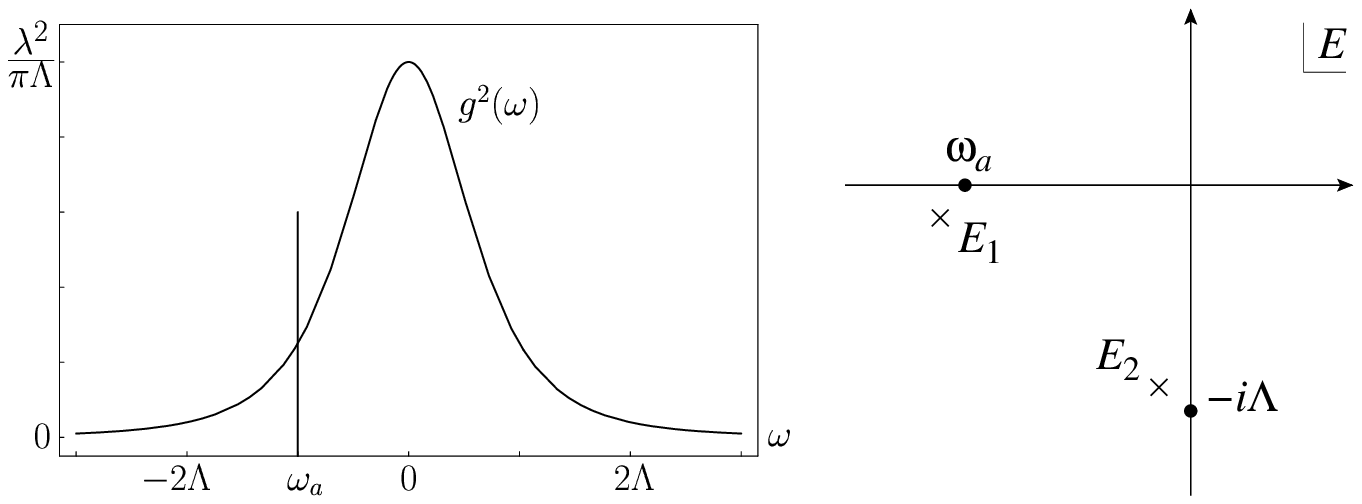,width=\textwidth}}
\caption{(a) Form factor $g^2(\omega)$ and initial state energy
$\omega_a$. (b) Poles of the propagator in the complex $E$-plane.}
\label{fig:formfact}
\end{figure}

We consider now a particular solvable model: let the form factor
in (\ref{eq:contham}) be Lorentzian
\andy{2formfact}
\beq
\label{eq:2formfact}
g(\omega)=\frac{\lambda}{\sqrt{\pi}}\sqrt{\frac{\Lambda}{\omega^{2}+\Lambda^{2}}}.
\eeq
This describes, for instance, an atom-field coupling in a cavity
with high finesse mirrors.\cite{Lang73} (Notice that the
Hamiltonian in this case is not lower bounded and we expect no
deviations from exponential behavior at very large
times.\cite{Khalfinlong}) In this case one easily obtains (for
$\Im E>0$)
\andy{2selfen}
\beq
\Sigma_a(E)=\frac{\lambda^2}{E+i\Lambda},
\label{eq:2selfen}
\eeq
whence the propagator
\andy{2propag}
\beq
\label{eq:2propag}
G_a(E)=\frac{E+i\Lambda}{(E-\omega_a)(E+i\Lambda)-\lambda^2}
\eeq
has two poles in the lower half energy plane (see Fig.\
\ref{fig:formfact}). Their values are
\andy{2poles}
\beq
\label{eq:2poles}
E_1=\omega_a+\delta\omega_a-i\frac{\gamma}{2}, \quad
E_2=-\delta\omega_a-i\left(\Lambda-\frac{\gamma}{2}\right),
\eeq
where
\andy{2poles1}
\beq
\left\{\begin{array}{c}
 \delta\omega_a=-\frac{\omega_a}{2}+\frac{\omega_a}{2}
 \sqrt{\frac{\sqrt{\upsilon^4+4\omega_a^2\Lambda^2}+\upsilon^2}
 {2\omega_a^2}} \\
 \\
 \gamma=\Lambda-
 \sqrt{\frac{\sqrt{\upsilon^4+4\omega_a^2\Lambda^2}-\upsilon^2}
 {2}}
\end{array}\right. ,\quad
\mbox{with}\quad \upsilon^2=\omega_a^2+4\lambda^2-\Lambda^2 .
\label{eq:2poles1}
\eeq
(Notice that $\upsilon^2$ can be negative.)
The propagator and the survival amplitude read
\andy{2survprop}
\barr
G_a(E)&=&\frac{E_1+i\Lambda}{E_1-E_2} \frac{1}{E-E_1}
-\frac{E_2+i\Lambda}{E_1-E_2} \frac{1}{E-E_2} \nonumber\\
&=& \frac{1-\cR}{E-(\omega_a+\delta\omega_a)+i\gamma/2}+
\frac{\cR}{E+\delta\omega_a +i(\Lambda-\gamma/2)}
\label{eq:2survprop}
\earr
and
\andy{2survamp}
\barr
\As(t)
=(1-\cR) e^{-i(\omega_a+\delta\omega_a)t} e^{-\gamma t/2} + \cR
e^{i\delta\omega_a t} e^{-(\Lambda-\gamma/2) t} ,
\label{eq:2survamp}
\earr
respectively, where
\andy{renorma}
\beq
1-\cR=\mbox{Res}[G_a(E_1)]= \frac{1}{1-\Sigma_a^\prime(E_1)}
=\frac{\omega_a+\delta\omega_a+i(\Lambda-\gamma/2)}{\omega_a+2\delta\omega_a+i(\Lambda-\gamma)}
\label{eq:renorma}
\eeq
is the residue of the pole $E_1$ of the propagator. The survival
probability reads
\andy{2survprob}
\barr
P(t) =|\As(t)|^2 &=& \cZ \exp(-\gamma t) +2\mbox{Re}[\cR^* (1-\cR)
e^{-i(\omega_a+2\delta\omega_a)t}]\exp(-\Lambda t)
\nonumber\\& &
+ |\cR|^2\exp[-(2\Lambda-\gamma)t],
\label{eq:2survprob}
\earr
where $\cZ=|1-\cR|^2$ is the wave function renormalization
\andy{zz}
\beq
\label{eq:zz}
{\cal Z} =
\frac{(\omega_a+\delta\omega_a)^2+(\Lambda-\gamma/2)^2}{(\omega_a+2\delta\omega_a)^2+(\Lambda-\gamma)^2}\
.
\eeq
All the above formulas are exact. We now analyze some interesting
limits of the model investigated.

\subsection{Small coupling}

In the weak coupling limit $\lambda\ll\omega_a,\Lambda$, one
obtains from Eq.\ (\ref{eq:2poles1})
\andy{smallOmega}
\barr
\delta\omega_a&=& \frac{\lambda^2}{\omega_a^2+\Lambda^2}\omega_a +
\Ord(\lambda^4)={\rm P}\!\!\int
d\omega\frac{g^2(\omega)}{\omega_a-\omega}+\Ord(\lambda^4),
\nonumber
\\
\gamma&=&2\Lambda\frac{\lambda^2}{\omega_a^2+\Lambda^2} +
\Ord(\lambda^4)=2\pi g^2(\omega_a)+\Ord(\lambda^4)  .
\label{eq:smallOmega}
\earr
Notice that the latter formula is the Fermi Golden Rule and $E_1$
in (\ref{eq:2poles}) is the ``dominant" pole. Indeed, the second
exponential in Eq. (\ref{eq:2survamp}) is damped very quickly, on
a time scale $\Lambda^{-1}$ much faster than $\gamma^{-1}$,
whence, after a short initial quadratic (Zeno) region of duration
$\Lambda^{-1}$, the decay becomes purely exponential with decay
rate $\gamma$. Notice that the corrections are of order
$\lambda^2$
\andy{smOmR}
\beq
\cR=\frac{\lambda^2}{\omega_a^2+\Lambda^2}\frac{\omega_a-i\Lambda}{\omega_a+i\Lambda}
+\Ord(\lambda^4)
\label{eq:smOmR}
\eeq
and the Zeno time is $\tau_{\rm Z}=\lambda^{-1}\gg \Lambda^{-1}$,
i.e. the initial quadratic (Zeno) region is {\em much shorter}
than the Zeno time: in general, the Zeno time does {\em not} yield
a correct estimate of the duration of the Zeno
region.\cite{FP98,PIO,Antoniou} (Beware of many erroneous claims
in the literature!) The approximation $P(t)
\simeq 1 - t^2/\tau_{\rm Z}^2$  holds for times $t < \Lambda^{-1}
\ll
\tau_{\rm Z}$.

\subsection{Large bandwidth}
In the limit of large bandwidth $\Lambda\gg\omega_a,\lambda$, from
Eq. (\ref{eq:2poles1}) one gets
$\gamma=2\lambda^2/\Lambda+\Ord(\Lambda^{-2})$ and in order to
have a non trivial result with a finite decay rate, we let
\andy{flatbandlim}
\beq
\Lambda\to\infty, \quad \lambda\to\infty, \qquad \mbox{with} \quad
\frac{\lambda^2}{\Lambda}=\frac{\gamma}{2}=\mbox{const}.
\label{eq:flatbandlimit}
\eeq
In this limit the continuum has a flat band,
$g(\omega)=\sqrt{\gamma/2\pi}=$const, and we expect to recover a
purely exponential decay. Indeed, in this case one gets $\cR=0$
and $\delta\omega_a=0$, whence
\andy{poleprop}
\beq
G_a(E)=\frac{1}{E-\omega_a+i\gamma/2},
\label{eq:poleprop}
\eeq
so that the survival amplitude and probability read
\andy{purexp}
\beq
\As(t)=\exp\left(-i\omega_a t-\frac{\gamma}{2}t\right)  \quad
\mbox{and} \quad P(t)=\exp(-\gamma t).
\label{eq:purexp}
\eeq
In this case the propagator (\ref{eq:poleprop}) has only a simple
pole and the survival probability (\ref{eq:purexp}) is purely
exponential.

\subsection{Narrow bandwidth}
\label{sec-narrow}
 In the limit of narrow bandwidth
$\Lambda\ll\omega_a,\lambda$, the form factor becomes
\beq
g^2(\omega)=\lambda^2 \delta(\omega)
\label{eq:twolevform}
\eeq
and the continuum is ``concentrated" in $\omega=0$. Therefore the
continuum as a whole behaves as another discrete level and one
obtains Rabi oscillations between the initial state $\ket{a}$ in
(\ref{eq:contham}) and this ``collective" level (of energy
$\omega=0$). In fact one gets
\andy{parm0Lam}
\beq
\gamma=0,\quad\delta\omega_a=-\frac{\omega_a}{2}+\Omega, \quad
\cR=\frac{1}{2}\left(1-\frac{\omega_a}{2\Omega}\right),
\label{eq:parm0Lam}
\eeq
where
\andy{Rabimod}
\beq
\Omega=\sqrt{\lambda^2+\frac{\omega_a^2}{4}}
\label{eq:Rabimod}
\eeq
is the usual Rabi frequency of a two-level system with energy
difference $\omega_a (=\omega_a-0)$ and coupling $\lambda$. By
(\ref{eq:parm0Lam}) the survival amplitude and probability read
\andy{surv0Lam}
\barr
\As(t)&=&\frac{1}{2}\left(1+\frac{\omega_a}{2\Omega}\right)
e^{-i(\omega_a +\Omega)t}
+\frac{1}{2}\left(1-\frac{\omega_a}{2\Omega}\right) e^{-i(\omega_a
-\Omega)t}, \nonumber\\ P(t)&=& 1-\frac{\lambda^2}{\Omega^2}
 \sin^2\left(\frac{\Omega t}{2}\right).
\label{eq:surv0Lam}
\earr
If $\omega_a=0$, the survival probability\ (\ref{eq:surv0Lam})
oscillates between $1$ and $0$. On the other hand, if
$\omega_a\neq0$ the initial state never ``decays" completely.

Incidentally, notice that the Zeno time is still $\tau_{\rm
Z}=\lambda^{-1}$ and yields now a good estimate of the duration of
the Zeno region. This is, so to say, a ``coincidence" due to the
oscillatory features of the system.

\subsection{Strong coupling}

Another interesting case is that of strong coupling, $\lambda
\simeq \Lambda$. This is a typical case in which the strong
coupling provokes violent oscillations before the system reaches
the asymptotic regime. In the limit $\lambda\gg\Lambda,\omega_a$,
we get
\beq
\delta\omega_a= \lambda -\frac{\omega_a}{2}+\Ord(\lambda^{-1}), \quad
\gamma= -i\frac{\Lambda}{2}+\Ord(\lambda^{-1}), \quad
\cR=\frac{1}{2}-\frac{\omega_a+i\Lambda}{4\lambda}+\Ord(\lambda^{-3}),
\eeq
whence the survival amplitude reads
\beq
\As(t)\simeq \exp\left(-i\frac{\omega_a}{2} t-\frac{\Lambda}{2}
t\right)\left[\left(\frac{1}{2}+\frac{\omega_a+i\Lambda}{4\lambda}\right)e^{-i\lambda
t}+\left(\frac{1}{2}-\frac{\omega_a+i\Lambda}{4\lambda}\right)e^{i\lambda
t} \right],
\eeq
which yields fast oscillations of frequency $\lambda$ damped at a
rate $\Lambda\ll\lambda$.

\subsection{Two-pole reduction}
We now show that the two-pole model introduced in this section is
the first improvement, after the Weisskopf-Wigner pole, in the
approximation of a generic quantum field model. First note that,
according to the Weisskopf-Wigner approximation (\ref{eq:WW0}), an
exponential decay is obtained by considering a constant
self-energy function $\Sigma_a=-i\gamma/2$, i.e. a resolvent with
a single pole with negative imaginary part ($E_1$ in Figure
\ref{fig:formfact}). On the other hand, as we noted in Sec.\
\ref{sec-unstable}, the initial quadratic behavior of the survival
amplitude is governed by two effective poles of the resolvent,
which ultimately derive from the behavior\ (\ref{eq:Sigma(1/u)1})
of the self-energy function at infinity
\beq
\Sigma_a(E)\sim \frac{1}{\tau_{\rm Z}^2 E}, \qquad\mbox{for} \quad
E\to\infty.
\label{eq:2poleshort}
\eeq
If one wants to capture this short-time behavior while keeping the
exponential law at later times, and is not interested in the
long-time power-law deviations, one can proceed in the following
way. The requirement for having an exponential decay, with decay
rate $\gamma$ for $t\to\infty$, translates into the behavior of
the self-energy function for $E\to 0$, namely in the requirement
of having a Weisskopf-Wigner constant self-energy function with
negative imaginary part
\beq
\Sigma_a(0)=-i b. \quad (b>0)
\label{eq:2polelarge}
\eeq
The simplest form of the self-energy function satisfying both
requirements\ (\ref{eq:2poleshort}) and\ (\ref{eq:2polelarge}) is
\beq
\Sigma_a(E)=\frac{1}{\tau_{\rm Z}^2E+i/b}=\frac{1/\tau_{\rm Z}^2
}{E+i/b\tau_Z^2}
\eeq
By letting $\tau_{\rm Z}=1/\lambda$ and $1/b\tau_Z^2=\Lambda$,
this becomes exactly the self-energy function of the two-pole
model\ (\ref{eq:2selfen}). Therefore the two-pole model is the
simplest approximation which yields the short time quadratic
behavior together with the long time exponential one.

We call the technique outlined in this subsection ``two-pole
reduction." It is useful if one wants to get a first idea of the
temporal behavior of a quantum field, keeping information on the
lifetime (Fermi golden rule), but also on the short-time Zeno
region.

Note that the  process outlined above can be iterated to find
better approximations of the real self-energy function
$\Sigma_a(E)$ by adding other poles and/or zeros. But notice also
that this approach does not yield the inverse power-law tail.
Indeed the latter is essentially due to the nonanalytic behavior
of the self-energy function at the branching point, a feature that
cannot be captured by an olomorphic function.

\section{Modification of the Lee Hamiltonian }
We now introduce an interesting modification of the Lee
Hamiltonian\ (\ref{eq:contham}), which enables us to look at the
Zeno region from a different perspective. The Hamiltonian
(\ref{eq:contham}) describes the decay of a discrete state
$\ket{a}$ into a continuum of states $\ket{\omega}$ with a given
form factor $g(\omega)$. According to Eqs.\ (\ref{eq:tzoff}) and
(\ref{eq:contham}), the Zeno time is related to the integral of
the squared form factor by the simple relation
\beq
\label{eq:tauZform2}
\frac{1}{\tau_{\rm Z}^2}=\int d\omega\; g^2(\omega).
\eeq
On the other hand, for a two-level system $\{\ket{a},\ket{b}\}$
with Hamiltonian
\beq
\label{eq:HtauZ}
H=\lambda (\ket{a}\bra{b}+\ket{b}\bra{a}), \quad \tau_{\rm
Z}=1/\lambda
\eeq
the Zeno time $\tau_{\rm Z}$ is just the inverse off-diagonal
element $\lambda$ of the Hamiltonian [and, of course, this is in
agreement with Eq.\ (\ref{eq:tauZform2}), as shown by Eq.\
(\ref{eq:twolevform})]. One has therefore a simple system in which
the Zeno time is manifest in the Hamiltonian itself. We seek now
an equivalent decay model, that shares with the two-level model
this nice property. To this end, let us add a new ``intermediate"
discrete state $\ket{b}$ to the Lee model. Consider then the Rabi
oscillation $\lambda$ of the two-level system $\ket{a}$, $\ket{b}$
and let the initial state $\ket{a}$ decay only {\em through} state
$\ket{b}$, i.e.\ couple $\ket{b}$ to a continuum with form factor
$g_b(\omega)$. In other words, the Hamiltonian\ (\ref{eq:contham})
is substituted by the following one
\andy{conthamb}
\barr
H &=& \omega_a\ket{a}\bra{a} +\omega_b\ket{b}\bra{b} +\int
d\omega\;\omega \ket{\omega} \bra{\omega}
\nonumber \\
& & +
\lambda\left(\ket{a}\bra{b}+\ket{b}\bra{a}\right)+ \int
d\omega\;g_b(\omega) ( \ket{b} \bra{\omega}+ \ket{\omega}
\bra{b}) .
\label{eq:conthamb}
\earr
We require that this Hamiltonian is equivalent to the original one
in describing the decay of the initial state $\ket{a}$. To this
end, notice that the part of Hamiltonian describing the decay of
state $\ket{b}$ (and neglecting the coupling with $\ket{a}$) is
just a Lee Hamiltonian and gives
\beq
G_b(E)=\frac{1}{E-\omega_b-\Sigma_b(E)}, \qquad \Sigma_b(E)=\int
d\omega\; \frac{g_b^2(\omega)}{E-\omega}.
\eeq
On the other hand, state $\ket{a}$ couples only to state $\ket{b}$
with a coupling $\lambda$. Therefore the evolution of state
$\ket{a}$ is just a Rabi oscillation between state $\ket{b}$
dressed by the continuum $\ket{\omega}$ and state $\ket{a}$,
namely
\beq
G_a=G_{a}^0+G_{a}^0 \lambda G_b \lambda G_a\ ,
\eeq
whence
\beq
G_a(E)=\frac{1}{E-\omega_a-\lambda^2 G_b(E)}\ .
\eeq
Therefore, in the modified model, the self-energy function of the
initial state $\ket{a}$ is nothing but the coupling $\lambda^2$
times the dressed propagator $G_b(E)$
\beq
\Sigma_a(E)=\lambda^2
G_b(E)=\frac{\lambda^2}{E-\omega_b-\Sigma_b(E)} .
\label{eq:SigaSigb}
\eeq
Equation\ (\ref{eq:SigaSigb}) is the equivalence relation sought.
One has to choose the auxiliary form factor $g_b(\omega)$ in Eq.\
(\ref{eq:conthamb}) as a function of the original one $g(\omega)$,
in order to satisfy this relation and get an equivalent
description of the decay. Our interest in this equivalence is due
to the asymptotic behavior of formula (\ref{eq:SigaSigb})
\beq
\Sigma_a(E)\sim\frac{\lambda^2}{E}=\frac{1}{\tau_{\rm Z}^2 E},
\qquad\mbox{for}\quad E\to\infty\ ,
\eeq
which displays the relation between the coupling $\lambda$ and the
Zeno time $\tau_{\rm Z}$. Thus the Hamiltonian (\ref{eq:conthamb})
explicitly reads
\andy{conthamb1}
\barr
H &=& \omega_a\ket{a}\bra{a} +\omega_b\ket{b}\bra{b} +
\frac{1}{\tau_{\rm Z}}\left(\ket{a}\bra{b}+\ket{b}\bra{a}\right)
\nonumber \\
& & +\int d\omega\;\omega \ket{\omega} \bra{\omega} + \int
d\omega\;g_b(\omega) ( \ket{b} \bra{\omega}+
\ket{\omega}
\bra{b}) .
\label{eq:conthamb1}
\earr
In the equivalent model, therefore, the initial quadratic behavior
is singled out from the remaining part of the decay: the Zeno
region, i.e.\ the first oscillation, is nothing but the initial
unperturbed Rabi oscillation between states $\ket{a}$ and
$\ket{b}$ (which initially ``represents" the original continuum as
a whole). After the initial stage of the decay, the coupling
$g_b(\omega)$ between $\ket{b}$ and $\ket{\omega}$ (namely the
details of the original continuum) comes into play and modifies
the initial Rabi oscillation with its characteristic time scale.
This explains from a different perspective the difference, already
stressed in previous sections, between the Zeno time and the
duration of the initial quadratic region.

\begin{figure}[t]
\centerline{\epsfig{file=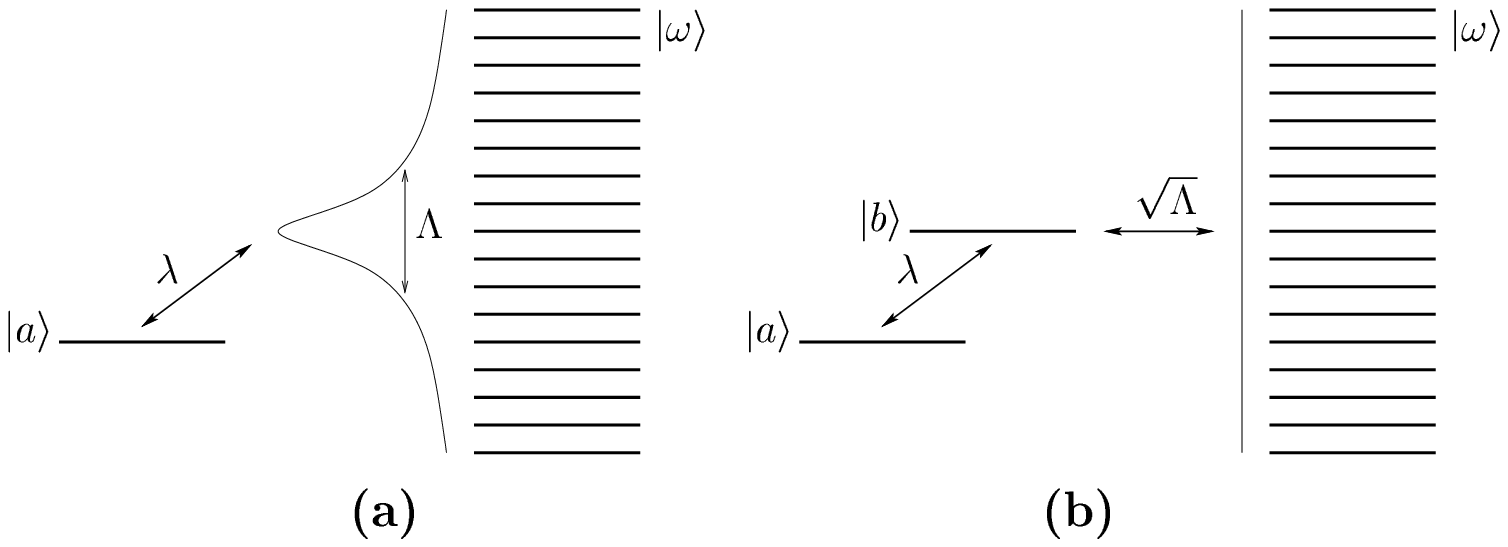,width=\textwidth}}
\caption{The decay of state $\ket{a}$ into a Lorentzian continuum $\ket{\omega}$
(a) is exactly equivalent to a Rabi coupling of $\ket{a}$ with a
state $\ket{b}$ that in turn exponentially decays into a flat
continuum $\ket{\omega}$ (b).}
\label{fig:equivalent}
\end{figure}

As an example, we recover the self-energy function
(\ref{eq:2selfen}) of the two-pole model, by requiring that
$\Sigma_b$ be constant
\beq
\Sigma_b(E)=-\omega_b-i\Lambda\ ,
\eeq
which implies
\beq
g_b(\omega)=\sqrt{\Lambda/\pi} \qquad\mbox{and} \quad
\omega_b=0.
\eeq
In other words, the auxiliary state $\ket{b}$ is placed at the
mean energy of the original continuum $g(\omega)$ and decays into
a flat-band continuum with decay rate $\gamma_b=2\Lambda$: the
decay into a Lorentzian continuum [Fig.\ \ref{fig:equivalent}(a)]
is exactly equivalent to a Rabi coupling with a level that in turn
exponentially decays into a flat continuum [Fig.\
\ref{fig:equivalent}(b)]:
\barr
H&=&\omega_a \ket{a}\bra{a} +
\frac{1}{\tau_{\mathrm{Z}}} \left(\ket{a}\bra{b}+\ket{b}\bra{a}\right)
+\int d\omega\;\omega \ket{\omega}
\bra{\omega}
\nonumber\\
& & +\sqrt{\frac{\Lambda}{\pi}} \int d\omega\; ( \ket{b}
\bra{\omega}+
\ket{\omega}
\bra{b}) .
\earr
Moreover, if one restricts one's attention to the subspace spanned
by $\{\ket{a},\ket{b}\}$, it is easy to show\cite{PIO} that this
Hermitian Hamiltonian reduces to the effective non-Hermitian one
\barr
H&=&\omega_a \ket{a}\bra{a} +
\lambda \left(\ket{a}\bra{b}+\ket{b}\bra{a}\right)
-i\Lambda \ket{b}\bra{b}\nonumber\\
& =&\left(\begin{array}{cc}
  \omega_a & \lambda \\
  \lambda & -i\Lambda
\end{array}\right)
=\left(\begin{array}{cc}
  \omega_a & \tau_{\mathrm{Z}}^{-1} \\
  \tau_{\mathrm{Z}}^{-1} & -i \left(\frac{\gamma}{2}+\frac{2}{\tau_{\mathrm{Z}}^2
  \gamma}\frac{1}{1+4\delta\omega_a^2/\gamma^2}\right)
\end{array}\right)
\earr
Therefore, if one is interested only in the decay of the initial
state $\ket{a}$, the study of the two-pole model reduces to the
study of this simple non-Hermitian 2-dimensional matrix. One can
reexamine all the results of previous sections, just by looking at
this matrix. We will not elaborate on this here.

A final comment is now in order. One can draw a clear picture of
the two-pole reduction, discussed in the previous section, just by
looking at the construction of the equivalent model. The first
approximation of a real decay, the Weisskopf-Wigner approximation,
is represented by the simple exponential decay of level $\ket{b}$
with its time scale $\gamma_b^{-1}$. The two-pole approximation
superimposes an oscillating dynamics with time scale
$\lambda^{-1}$ to the latter, yielding the initial Zeno region. By
complicating the model with the addition of other dynamical
elements with their characteristic scales, one can construct a
better approximation of the real decay law.

\section{Zeno--Heraclitus transition}
We will now study the Zeno--inverse Zeno transition in greater
detail, by making use of a quantum field theoretical framework,
and discuss the primary role played by the form factors of the
interaction. The reader should refer to the discussion of Secs.\
\ref{sec-dpw}-\ref{sec-QZEinv},
where we introduced the effective decay rate
\andy{effgamma09}
\beq
\gamma_{\rm eff}(\tau) \equiv -\frac{1}{\tau}\log
P(\tau)=-\frac{2}{\tau}\log |\As(\tau)|=-\frac{2}{\tau}
\Re\Big[\log\As(\tau)\Big],
 \label{eq:effgamma09}
\eeq
which is a linear function of $\tau$ for sufficiently small values
of $\tau$ (inside the Zeno region)
\andy{effgamma9}
\beq
\gamma_{\rm eff}(\tau) \sim \frac{\tau}{\tau_{\rm Z}^2}, \qquad
\mbox{for}\quad \tau \lesssim 1/\Lambda ,
 \label{eq:effgamma9}
\eeq
and becomes, with excellent approximation, a constant equal to the
natural decay rate at intermediate times
\andy{effgammalim9}
\beq
\gamma_{\rm eff}(\tau) \simeq  \gamma \qquad \mbox{for}\quad \tau
\gg 1/\Lambda.
 \label{eq:effgammalim9}
\eeq
The transition between Zeno and Heraclitus occurs at the
geometrical intersection $\tau^*$ between the curves $P(t)$ and
$e^{-\gamma t}$, solution of the equation
\andy{tstardef9}
\beq
\gamma_{\rm eff}(\tau^*)=\gamma\ ,
\label{eq:tstardef9}
\eeq
as shown in Fig.\ \ref{fig:gtau}.

\begin{figure}[t]
\centerline{\epsfig{file=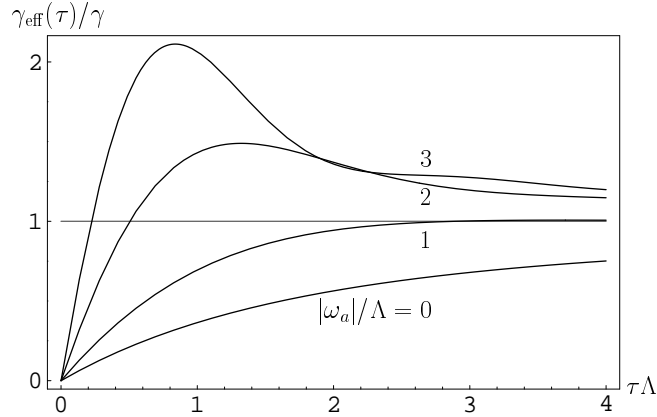,height=5.5cm}}
\caption{Effective decay
 rate $\gamma_{\rm eff}(\tau)$ for the two-pole model
(\ref{eq:xlor}), for $\lambda=0.1$ and different values of the
ratio $|\omega_a|/\Lambda$ (indicated). The horizontal line shows
the ``natural" decay rate $\gamma$: its intersection with
$\gamma_{\rm eff}(\tau)$ yields the solution $\tau^*$ of Eq.\
(\ref{eq:tstardef9}). The asymptotic value of all curves is
$\gamma$, as expected. A Zeno (inverse Zeno) effect is obtained
for $\tau<\tau^*$ ($\tau>\tau^*$). Notice the presence of a linear
region for small values of $\tau$ and observe that $\tau^*$ does
not belong to such linear region as the ratio $|\omega_a|/\Lambda$
decreases. Above a certain threshold, given by Eq.\
(\ref{eq:anticond1}) in the weak coupling limit of the model (and
in general by the condition $\cZ=1$), Eq.\ (\ref{eq:tstardef9})
has no finite solutions: only a Zeno effect is realizable in such
a case.}
\label{fig:gammat}
\end{figure}

Let us corroborate these general findings by considering for
example the two-pole model studied in detail in Sec.\
\ref{sec-two-pole}, whose survival amplitude is given by Eq.\
(\ref{eq:2survamp})
\andy{xlor}
\barr
\label{eq:xlor}
\As(t)&=&
\frac{\omega_a+\delta\omega_a+i(\Lambda-\gamma/2)}{\omega_a+2\delta\omega_a+i(\Lambda-\gamma)}
e^{-i(\omega_a+\delta\omega_a)t} e^{-\gamma t/2}
\nonumber \\
& &
+\frac{\delta\omega_a-i\gamma/2}{\omega_a+2\delta\omega_a+i(\Lambda-\gamma)}
e^{i\delta\omega_a t} e^{-(\Lambda-\gamma/2) t},
\earr
with $\delta\omega_a$ and $\gamma$ given by Eq.\
(\ref{eq:2poles1}).
 By plugging (\ref{eq:xlor}) into
(\ref{eq:effgamma09}) one obtains the effective decay rate, whose
behavior is displayed in Fig.~\ref{fig:gammat} for different
values of the ratio $|\omega_a|/\Lambda$. These curves show that
for large values of $|\omega_a|$ (in units $\Lambda$) there is
indeed a transition from a Zeno to an inverse Zeno (Heraclitus)
behavior: such a transition occurs at $\tau=\tau^*$, solution of
Eq.\ (\ref{eq:tstardef9}). However, for small values of
$|\omega_a|$, such a solution ceases to exist.

The determination of the critical value of $|\omega_a|$ for which
the Zeno--inverse Zeno transition ceases to take place discloses
an interesting aspect of this issue. The problem can be discussed
in general, but for the sake of simplicity we consider the weak
coupling limit (small $\lambda$) considered in Eqs.\
(\ref{eq:smallOmega})-(\ref{eq:smOmR}). According to the
geometrical theorem proved in Sec.\ \ref{sec-QZEinv}, a sufficient
condition for the system to exhibit an Zeno--inverse Zeno
transition is that the wave function renormalization $\cZ<1$. In
our case, by making use of Eq.\ (\ref{eq:smOmR}),  this condition
reads
\andy{anticond}
\barr
\label{eq:antocond}
\cZ&=&|1-\cR|^2=1-2\frac{\lambda^2}{\omega_a^2+\Lambda^2}\mbox{Re}
\left[\frac{\omega_{0}-i\Lambda}{\omega_a+i\Lambda}\right]
+\Ord(\lambda^4)
\nonumber \\
&=&1-2\frac{\lambda^2}{\omega_a^2+\Lambda^2}
\frac{\omega_a^2-\Lambda^2}{\omega_a^2+\Lambda^2}+\Ord(\lambda^4)<1,
\earr
namely
\andy{anticond1}
\beq
\label{eq:anticond1}
\omega_a^2>\Lambda^2+\Ord(\lambda^2).
\eeq
The meaning of this relation is the following: a sufficient
condition to obtain a Zeno--inverse Zeno transition is that the
energy of the decaying state be placed asymmetrically with respect
to the peak of the form factor (bandwidth) (see Fig.\
\ref{fig:formfact}). If, on the other hand, $\omega_a \simeq 0$
(center of the bandwidth), no transition time $\tau^*$ exists (see
Fig.\ \ref{fig:gammat}) and only a QZE is possible: this is the
case analyzed in Fig.\ \ref{fig:Zren}(a).

There is more: Equation (\ref{eq:xlor}) yields a time scale.
Indeed, from the definitions of the quantities in\
(\ref{eq:2poles1}) one gets $\gamma/2 < \Lambda - \gamma/2$, so
that the second exponential in (\ref{eq:xlor}) vanishes more
quickly than the first one. (The two time scales become comparable
only in the strong coupling regime: $\gamma\to\Lambda$ as
$\lambda\to\infty$.) If the coupling is weak, since
$\gamma=\mbox{O}(\lambda^2)$, the second term is very rapidly
damped so that, after a short initial quadratic region of duration
$\Lambda^{-1}$, the decay becomes purely exponential with decay
rate $\gamma$. For $\tau \lesssim 1/\Lambda$ (which is, by
definition, the duration of the quadratic Zeno region), we can use
the linear approximation\ (\ref{eq:effgamma9}). When it applies up
to the intersection (i.e., $|\omega_a|\gg \Lambda$) one gets
 \andy{jump}
\beq
\label{eq:jump}
 \tau^* \simeq \gamma \tau_{\rm Z}^2.
\eeq
When $\omega_a$ gets closer to the peak of the form factor, the
linear approximation does not hold and the r.h.s.\ of the above
expression yields only a lower bound to the transition time
$\tau^*$. In this case the solution $\tau^*$ of Eq.\
(\ref{eq:tstardef9}) becomes larger than the approximation
(\ref{eq:jump}), eventually going to infinity when condition
(\ref{eq:anticond1}) is no longer valid. In such a case, only a
QZE is possible and no IZE is attainable. This is in contrast with
recent claims.\cite{KKclaim}

The conclusions obtained for the two-pole model (\ref{eq:xlor})
are of general validity.  In general, in the Lee Hamiltonian\
(\ref{eq:contham}), for any $g(\omega)$, we assume that
$\omega_a>\omega_g$ (the lower bound of the continuous spectrum),
in order to get an unstable system. The matrix elements of the
interaction Hamiltonian depend of course on the physical model
considered. However, for physically relevant situations, the
interaction smoothly vanishes for small values of
$\omega-\omega_g$ and quickly drops to zero for $\omega
> \Lambda$, a frequency cutoff related to the size of the decaying
system and the characteristics of the environment. This is true
both for cavities, as well as for typical EM decay processes in
vacuum, where the bandwidth $\Lambda \simeq
10^{14}-10^{18}$s$^{-1}$ is given by an inverse characteristic
length\cite{heraclitus,PIO} (say, of the order of Bohr radius) and
is much larger than the natural decay rate $\gamma\simeq
10^{7}-10^{9}$s$^{-1}$.

For roughly bell-shaped form factors all the conclusions drawn for
the Lorentzian model remain valid. The main role is played by the
ratio $\omega_{ag}/\Lambda$, where $\omega_{ag}=\omega_a-\omega_g$
is the available energy. In general, the asymmetry condition
$\omega_{ag}<\Lambda$ is satisfied if the energy $\omega_a$ of the
unstable state is sufficiently close to the threshold. In fact,
from Eq.\ (\ref{eq:tauZform2}) one has
\beq
\frac{1}{\tau_{\rm Z}^2}=\int d\omega\;
g^2(\omega)=g^2(\bar\omega) \Lambda,
\eeq
where $\bar\omega$ is defined by this relation and is of order
$\omega_{\rm max}$, the energy at which $g(\omega)$ takes the
maximum value. For $\omega_a$  sufficiently close to the threshold
$\omega_g$ one has $g(\bar\omega)\gg g(\omega_a)$, the time scale
$\gamma\tau_{\rm Z}^2$ is well within the short-time regime,
namely
\beq
\gamma \tau_{\rm Z}^2=\frac{2\pi
g^2(\omega_a)}{g^2(\bar\omega)\Lambda}\ll \frac{1}{\Lambda},
\eeq
where the Fermi golden rule $\gamma = 2\pi g^2(\omega_a)$ has been
used, and therefore the estimate (\ref{eq:jump}) is valid.

On the other hand, for a system such that $\omega_{ag}
\simeq\Lambda$ (or, better, $\omega_a \simeq \;$center of the
bandwidth), $\tau^*$ does not necessarily exist and usually {\it
only} a Zeno effect can occur. In this context, it is useful and
interesting to remember that, as shown in Sec.\ \ref{sec-narrow},
the Lorentzian form factor (\ref{eq:2formfact}) yields, in the
limit $g^2(\omega)=\lambda^2\delta(\omega-\omega_a)$, the physics
of a two level system. This is also true in the general case, for
a roughly symmetric form factor, when the bandwidth $\Lambda\to
0$. In such a case, if $\omega_a=0$ (energy of the initial state
at the center of the form factor), the survival probability
oscillates between $1$ and $0$ and only a QZE is possible. On the
other hand, if $\omega_a\neq0$ (initial state energy strongly
asymmetric with respect to the form factor of ``width"
$\Lambda=0$) the initial state never decays completely. By
measuring the system, the survival probability will vanish
exponentially, independently of the strength of observation,
whence only an IZE is possible.

If one consider the large bandwidth limit of the two-pole model,
which is equivalent to a Weisskopf-Wigner approximation, the
propagator (\ref{eq:poleprop}) has only a simple pole and the
survival probability (\ref{eq:purexp}) is purely exponential.
Therefore measurements cannot modify the free behavior. Indeed,
the conditions for the occurrence of Zeno effects are always
ascribable to the presence of an initial non-exponential behavior
of the survival probability, which is caused by a propagator with
a richer structure than a simple pole in the complex energy plane.

\section{Continuous measurements}

Most of our examples dealt with instantaneous measurements,
according to the Copenhagen prescription. Our starting point was
indeed Eq.\ (\ref{eq:survN0}). However, it is always possible to
mimic the effect of pulsed measurements in terms of the coupling
to a suitable system, performing a continuous measurement process.
This issue has been discussed in other papers,\cite{PIO,Napoli} so
let us only give here an example. Let us add to (\ref{eq:contham})
the following Hamiltonian
\andy{omomprime}
\beq
H_{\rm meas}(\Gamma) = \sqrt{\frac{\Gamma}{2\pi}} \int d\omega
d\omega'\;( \ket{\omega} \bra{\omega, \omega'}+
\ket{\omega,\omega'} \bra{\omega}) + \int
d\omega'\; \omega' \ket{\omega'}\bra{\omega'}:
\label{eq:omomprime}
\eeq
as soon as a photon is emitted, it is coupled to another boson of
frequency $\omega'$ (notice that the coupling has no form factor).
One can show that the dynamics of the Hamiltonian
(\ref{eq:contham}) and (\ref{eq:omomprime}), in the relevant
subspace, is generated by
\andy{conthamsub}
\beq
H = \omega_a\ket{a}\bra{a} +\int d\omega\;(\omega -i \Gamma/2)
\ket{\omega} \bra{\omega} +\int d\omega\;g(\omega) ( \ket{a}
\bra{\omega}+ \ket{\omega} \bra{a}) ,
\label{eq:conthamsub}
\eeq
and an effective continuous observation on the system is obtained
by increasing $\Gamma$. Indeed, it is easy to see that the only
effect due to $\Gamma$ in Eq.\ (\ref{eq:conthamsub}) is the
substitution of $\Sigma_a(E)$ with $\Sigma_a(E+i\Gamma/2)$ in Eq.\
(\ref{eq:propag}), namely,
\andy{propagG}
\beq
\label{eq:propagG}
G_a(E)=\frac{1}{E-\omega_a-\Sigma_a(E+i\Gamma/2)}.
\eeq
For large values of $\Gamma$, i.e., for a very quick response of
the apparatus, the self-energy  function (\ref{eq:Sigma(1/u)1})
has the asymptotic behavior
\andy{selfenlargeG}
\beq
\label{eq:selfenlargeG}
\Sigma_a\left(E+i\frac{\Gamma}{2}\right)
\sim -i\frac{2}{\Gamma}\int d\omega\;
g^2(\omega)=-i\frac{2}{\Gamma \tau_{\rm Z}^2},\quad\mbox{for}\quad
\Gamma\to\infty .
\eeq
[Notice that $\Gamma \to \infty$ in (\ref{eq:selfenlargeG}) means
$\Gamma \gg \Lambda$, the frequency cutoff of the form factor.] In
this case the propagator (\ref{eq:propagG}) reads
\andy{propaglargeG}
\beq
\label{eq:propaglargeG}
G_a(E)\sim\frac{1}{E-\omega_a+i\gamma_{\rm eff}(\Gamma)/2},
\qquad\mbox{for}\quad \Gamma\to\infty
\eeq
and the survival probability decays with the effective exponential
rate (valid for $\Gamma\gg\Lambda$)
\andy{effgammalargeG}
\beq
\gamma_{\rm eff}(\Gamma) \sim \frac{4}{\tau_{\rm Z}^2 \Gamma}.
\label{eq:effgammalargeG}
\eeq
Notice the similarity of this result with (\ref{eq:lingammaeff}):
$\Gamma$, the strength of the coupling to the (continuously)
measuring system, plays the same role as $\tau^{-1}$, the
frequency of measurements in the pulsed formulation. This is a
general result.\cite{PIO,Schulman98} More to this, we have here a
scale for the validity of the linear approximation
(\ref{eq:effgammalargeG}) for $\gamma_{\rm eff}$: the linear term
in the asymptotic expansion (\ref{eq:selfenlargeG}) approximates
well the self-energy function only for values of $\Gamma$ that are
larger than the bandwidth $\Lambda$. For smaller values of
$\Gamma$ one has to take into account the nonlinearities arising
from the successive terms in the expansion.

Note that the flat-band case\ (\ref{eq:poleprop}), yielding a
purely exponential decay, is also unaffected by a continuous
measurement. Indeed in that case $\Sigma_a(E)=-i\gamma/2$ is a
constant independent of $E$, whence
$\Sigma_a(E+i\Gamma/2)=\Sigma_a(E)$ is independent of $\Gamma$.
The same happens if one considers the Weisskopf-Wigner
approximation\ (\ref{eq:WW0}): in this case one neglects the whole
structure of the propagator in the complex energy plane and
retains only the dominant pole near the real axis. This yields, as
we have seen, a self-energy function which does not depends on
energy and a purely exponential decay (without any deviations),
that cannot be modified by any observations.

\section{Conclusions}
\label{sec-conc}
\andy{sec-conc}

The form factors of the interaction Hamiltonian play a fundamental
role when the quantum system is ``unstable," not only because of
the very formulation of the Fermi golden rule, but also because
they may govern the transition from a Zeno to an inverse Zeno
(Heraclitus) regime. The inverse quantum Zeno effect has
interesting applications and turns out to be relevant also in the
context of quantum chaos and Anderson localization.\cite{chaos}

Although the usual formulation of QZE in terms of repeated
``pulsed" measurements {\em \`a la} von Neumann is a very
effective one and motivated quite a few theoretical proposals and
experiments, we cannot help feeling that the use of continuous
measurements (coupling with an external apparatus that gets
entangled with the system) is advantageous.

Both quantum Zeno and inverse quantum Zeno effects have been
experimentally confirmed. It is probably time to refrain from
academic discussions and look for possible applications.

\section*{Acknowledgments}
We thank Hiromichi Nakazato, Antonello Scardicchio and Larry
Schulman for interesting discussions.

\end{document}